\documentclass{appolb}
\usepackage{epsfig}
\usepackage{amssymb}
\def\la{\langle}
\def\ra{\rangle}
\def\kav{\la k_T\ra}
\def\k2av{\la k_T^2\ra}
\def\beq{\begin{equation}}
\def\eeq{\end{equation}}
\def\be{\begin{eqnarray}}
\def\ee{\end{eqnarray}}
\def\hs{\hat{s}}
\def\htm{\hat{t}}
\def\hu{\hat{u}}
\newcommand{\f}[2]{\frac{#1}{#2}}
\newcommand{\dd}{ {\textrm d}}

\begin{document}
\title{Differences in high $p_t$ meson production between CERN SPS
and RHIC heavy ion collisions%
%Perturbative QCD in High Energy Heavy Ion Collisions%
\thanks{Presented at Zakopane 2001 School, Zakopane, 2001 June}%
}
\author{G\'abor Papp$^1$, P\'eter L\'evai$^2$, Gergely G. Barnaf\"oldi$^2$,
Yi Zhang$^3$ and George Fai$^3$%
\address{$^1$ HAS Research Group for Theoretical Physics, P.O. Box 32,
Budapest 1518, Hungary  \\
$2$ KFKI Research Institute for Particle and Nuclear Physics,
P.O. Box 49, Budapest 1525, Hungary \\
$^3$ Center for Nuclear Research, Department of Physics,
         Kent State University, Kent, OH 44242}
}
\maketitle
\begin{abstract}
In this talk we present a perturbative QCD improved parton model 
calculation for
light meson production in high energy heavy ion collisions. In order
to describe the experimental data properly, one needs to augment the standard
pQCD model by the transverse momentum distribution of partons ("intrinsic 
$k_T$"). Proton-nucleus data indicate the presence of nuclear shadowing
and multiscattering effects. 
Further corrections are needed in nucleus-nucleus collisions to
explain the observed reduction of the cross section. We introduce the idea of
proton dissociation and compare our calculations with the SPS and RHIC 
experimental data.
\end{abstract}
  
\section{Introduction}
During the last decade, as the bombarding energy increased, nuclear
collision data have become available for high transverse momentum particle 
production. Perturbative quantum chromodynamics (pQCD) is believed to be
applicable in this regime. In this talk we present a pQCD based parton
model and test it in proton-proton ($pp$), proton-nucleus ($pA$), and
nucleus-nucleus ($AA$) collisions from CERN to Tevatron energies. Our
main goal is to understand first the elementary processes, like particle
production in $pp$ reactions, then learn about the mechanism of the nuclear
enhancement and finally, based on these studies, make ``predictions'' for
$AA$ collisions. Confronting our result to the available experimental data
we may look for any new phenomena, which cannot be described within the
original pQCD based parton model.

Expectations about what may happen at high energies or
large colliding systems include the formation of a quark gluon plasma (QGP).
In case QGP was formed in the
reaction, the outgoing jets would suffer an energy loss due to collisions
in the opaque plasma and a suppression in particle production would result.
This jet quenching~\cite{jetq} is expected to be prominent in high bombarding 
energy central heavy in collisions. Another effect is due to the fragility of
the proton, indicated by the fact that after a momentum transfer of 
$\sim 1-1.2 \ GeV$ it is blown to pieces~\cite{doksh01}. This proton
dissociation is also expected to happen at higher bombarding energies
and be most visible in central heavy ion collisions, where transverse momentum
may accumulate.

The talk is organized as follows: in the first section we discuss the basic
assumptions and formulae of a pQCD improved parton model, presenting it
at work in light meson production in $pp$ collisions.
Next, we present an analysis of the $pA$ collisions and
deduce some information on the nuclear enhancement 
(Cronin effect)~\cite{usbig}. 
Parameters fixed in this section will be used to study 
$AA$ collision at CERN SPS and RHIC energies.
Results deviate from experimental data for heavy ions. We present possible
explanations for this discrepancy and conclude that further experiments
are necessary to clarify the underlying physical picture.

\section{Light meson production in $pp$ collisions}

The invariant cross section for the production of hadron $h$ in a $pp$
collision  is described in the pQCD-improved parton model on the basis
of the factorization  theorem as a convolution~\cite{FF95}:  
\be
\label{hadX}
  E_{h}\f{\dd \sigma_h^{pp}}{\dd ^3p} &=
        \sum_{abcd} \int\! &\dd x_a \dd x_b \dd z_c\ f_{a/p}(x_a,Q^2)\
        f_{b/p}(x_b,Q^2)\ 
\f{\dd \sigma}{\dd \htm}(ab \to cd) \nonumber \\
   &&\times \frac{D_{h/c}(z_c, Q'^2)}{\pi z_c^2} \, \hs \, 
        \delta(\hs+\htm+\hu)\ \ \  , 
\ee
where  $f_{a/p}(x,Q^2)$ and  $f_{b/p}(x,Q^2)$  are the parton distribution
functions (PDFs) for the
colliding partons $a$ and $b$ in the interacting protons 
as functions of momentum fraction $x$, at scale $Q$,
$\dd \sigma/ \dd\htm$ is the hard scattering cross section of the
partonic subprocess $ab \to cd$, and the fragmentation function (FF), 
$D_{h/c}(z_c, Q'^2)$
gives the probability for parton $c$ to fragment into hadron $h$
with momentum fraction $z_c$ at scale $Q'$. We use the convention
that the parton-level Mandelstam variables are written with a `hat' (like 
$\htm$ above). We fix the scales as $Q = p_T/2$ and $Q' = p_T/(2z_c)$.

Such a model represents the ``hard'' physics and should not be pushed below
a scale $p_t \lesssim$ 1-2 GeV. In the following we restrict ourselves
to leading order (LO) pQCD, using the LO form of the partonic cross sections,
PDF's and FF's~\cite{KKP}.

It was noted as soon as pQCD calculations were applied to reproduce 
high-$p_T$ hadron production~\cite{owens87},
that this naive picture fails, especially at the lower end of the
$p_T$ range, $2\le p_T \le 6$ GeV. The partons participating
in meson production are bound inside nucleons and cannot be considered
to be in an infinite momentum frame. The concept of intrinsic 
transverse momentum was introduced~\cite{owens87} to take into account the
correction to the infinite momentum frame. A value of 
$\kav \sim 0.3-0.4$~GeV could be easily understood in terms of
the Heisenberg uncertainty relation for partons inside the proton.
However, a larger 
average transverse momentum of $\kav \sim 1$ GeV was extracted 
from jet-jet angular distributions (see e.g. \cite{E609}) and
explained theoretically as the effect of gluon rescattering inside
the nucleus~\cite{Wang9798}.

\begin{figure}
\epsfxsize=9cm \centerline{\epsffile{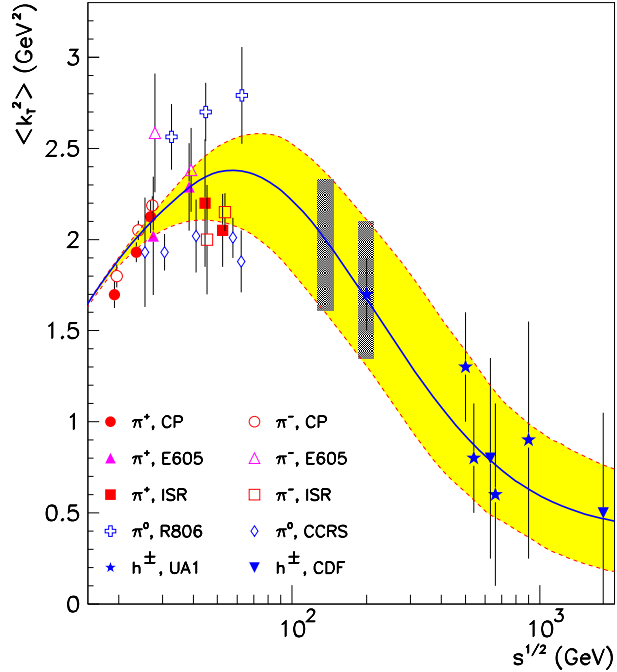}}
\vskip -0.05in
\caption[]{
 \label{figurekt2}
The best fit values of $\k2av$ in $p p \rightarrow \pi X$ 
\cite{antreasyan79,exps}
and $p \bar{p} \rightarrow h^\pm X$ \cite{UA1,CDF} reactions. 
Where large error bars would overlap at the same energy,
one of the points has been shifted slightly for better 
visibility. The band is drawn to guide the eye.
}
\end{figure}

Phenomenologically, the transverse momentum may be introduced by using a 
product assumption and extending each integral over the parton 
distribution functions to $k_T$-space~\cite{Wang9798},
\be
\label{ktbroad}
\dd x \ f_{a/p}(x,Q^2) \rightarrow \dd x 
\ \dd ^2\!k_T\ g({\vec k}_T) \  f_{a/p}(x,Q^2) \ ,
\ee
where $g({\vec k}_T)$ is the intrinsic transverse momentum distribution
of the relevant parton in the proton, and in this talk it is chosen
to be a Gaussian:
\beq
\label{kTgauss}
g({\vec k}_T) \ = \f{1}{\pi \la k^2_T \ra}
        e^{-{k^2_T}/{\la k^2_T \ra}}    \,\,\, .
\eeq

We made a systematic study of available $pp$ experiments producing
high $p_T$ pions and fitted the 2-dimensional width parameter 
$\la k_T^2 \ra$~\cite{usbig}. The best fit values are presented in 
Figure~\ref{figurekt2}. A similar plot can be obtained for
kaon production~\cite{usbig}.

\section{Proton-nucleus collisions}

Having fixed the intrinsic transverse momentum distribution in $pp$
collisions we turn now to $pA$ collisions and investigate the nuclear
enhancement (Cronin effect)~\cite{cronin}. It was found experimentally
that in the 2 GeV $<p_t<$ 5 GeV transverse momentum region there are
more particles produced than it was expected from a simple scaling of
$pp$ data. Within the present model this can be explained by an additional
term in the width of the intrinsic transverse momentum distribution,
which takes into account a broadening due to associated semihard inelastic
collisions:
\be
\label{ktbroadpA}
\k2av_{pA} = \k2av_{pp} + C \cdot h_{pA}(b) \ ,
\ee
where $\k2av_{pp}$ is the width of the transverse momentum distribution 
of partons in $pp$ collisions from the previous section, $h_{pA}(b)$ 
describes the number of {\it effective} nucleon-nucleon (NN) collisions 
at impact parameter $b$ which impart an average transverse momentum 
squared $C$. The coefficient $C$ is expected
to be approximately independent of $p_T$, of the target used,
and probably of beam energy (at least in the energy range studied 
in Ref.s \cite{antreasyan79,cronin}).

In $pA$
reactions, where one of the partons participating in the hard collision
originates in a nucleon with additional NN collisions, we will use the
$pp$ width from Fig. 1 for one of the colliding partons and the 
enhanced width~(\ref{ktbroadpA}) for the other. 
The effectivity function $h_{pA}(b)$ can be written in terms of the 
number of collisions suffered by the incoming proton in the target 
nucleus, $\nu_A(b) = \sigma_{NN} t_A(b)$, where $\sigma_{NN}$ is
the inelastic nucleon-nucleon cross section and 
$t_A(b)=\int \dd z \, \rho(b,z)$ is the nuclear thickness function.
Assuming that only the first $m-1$ semihard collisions preceeding the
hard collision contribute to the broadening of the 
width, we define the effectivity function as
\be
  h_{pA}(b) = \left\{ \begin{array}{cc}
                \nu_A(b)-1 & \nu_A(b) < m \\
                m-1 & \mbox{otherwise} \\
        \end{array} \right.\ .
\ee
The value $m=\infty$ corresponds to the scenario where all semihard
collisions contribute to the broadening. For realistic nuclei
$\nu_A(b)$ do not exceed the vale of 6, so we restrict ourselves to the region
$1<m<6$ and examine the dependence of the results on the possible choices
between these limits. 

According to the Glauber picture, the hard pion production cross section 
from $pA$ reactions can be written as an integral over impact parameter $b$:
\be
\label{pAX}
  E_{\pi}\f{\dd \sigma_{\pi}^{pA}}{ \dd ^3p} =
       \int \dd ^2b \,\, t_A(b)\,\, E_{\pi} \, \f{\dd \sigma_{\pi}^{pp}(\k2av_{pA},\k2av_{pp})}
{\dd ^3p}  
\,\,\, ,
\ee
with a further modification of the PDF's: in the nuclear 
environment ``shadowing'' effects~\cite{wang91,eskola99} 
modify the distribution
functions. Here we use an average, scale independent 
parameterization~\cite{wang91},
\be
\label{shadow}
f_{a/A}(x,Q^2) = S_{a/A}(x) \left[\frac{Z}{A} f_{a/p}(x,Q^2) + \left(1-\frac{Z}{A}\right)
  f_{a/n}(x,Q^2) \right]   \,\,\,\,  ,
\ee
where $f_{a/n}(x,Q^2)$ is the PDF for the neutron.

Confronting the calculations with experimental
data~\cite{antreasyan79} for Be, Ti and W targets at three different energies,
we obtain the best fit with $m=4$ and $C\approx$ 0.4 GeV$^2$~\cite{usbig}
and will use these values for $AA$ reactions in the next section.

\section{Nucleus-nucleus collisions}

Nucleus-nucleus collisions do not involve additional parameters in the
pQCD parton model with intrinsic $k_\perp$, both partons entering the
hard collision gain extra broadening of the width according 
to~(\ref{ktbroadpA}), i.e. depending on the number of nucleons within
the other nucleus in the channel swept by the particle. Thus,
\beq
\label{ABX}
  E_{\pi}\f{\dd \sigma_{\pi}^{AB}}{\dd ^3p} =
       \int \dd ^2b \, \dd ^2r \,\, t_A(r) \,\, t_B(|\vec b - \vec r|) \,
E_{\pi} \, \f{\dd \sigma_{\pi}^{pp}(\k2av_{pA},\k2av_{pB})}{\dd ^3p}  
\,\,\, ,
\eeq

In the following we calculate and compare to experimental data 
the pionic cross sections for CERN SPS reactions
with $\k2av_{pp}=$1.6 and 1.7~GeV$^2$ for 200 and 158 AGeV collisions,
respectively (see Fig.~\ref{figurekt2}), 
and with $m=4$, $C = 0.4$~GeV$^2$. Next, we investigate the
recent RHIC heavy ion collision at 130 GeV with $\k2av_{pp}=$2.0~GeV.

\subsection{CERN SPS energy}

Let us now confront the theoretical model~(\ref{ABX}) with the CERN SPS
experiments WA80~\cite{WA80} and WA98~\cite{WA98} for central collisions,
calculating the invariant cross section of pion production in
the 25\% most central $S+S$, 7.7\% most central $S+Au$, and 12.7\% most
central $Pb+Pb$ collisions within the experimental rapidity windows.

\begin{figure}
\epsfxsize=\textwidth \centerline{\epsffile{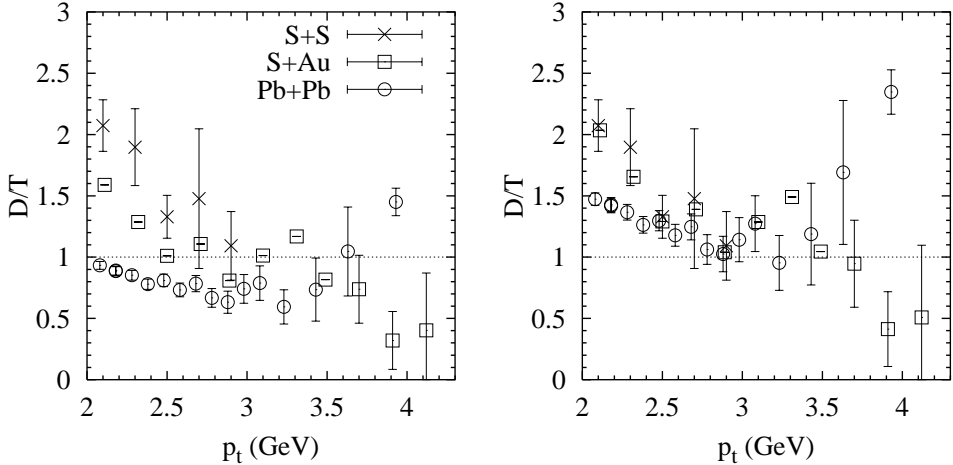}}
\vskip -0.05in
\caption[]{
 \label{figurenc}
Data/Theory ratio at CERN SPS $S+S$ (crosses), $S+Au$ (boxes) and
$Pb+Pb$ (circles) for high $p_t$ pion production reactions. Left: ratio 
without proton dissociation, right: ratio with proton dissociation after
4 collisions.
}
\end{figure}

The data over theory ratio ($D/T$) is presented in Figure~\ref{figurenc}
(left). For the lighter systems this ratio approaches 1 above
$p_t\gtrsim$ 2.5 GeV, while for the lead collisions we see that such a
model over-predicts the experimental values by 40\%. In the following we
speculate on the origin of this discrepancy.

A possible candidate for the reduction of the cross section in large
systems is the proton dissociation mentioned in the introduction.
Our $pA$ collision study showed
that each $pp$ inelastic collision adds $\sim$ 400 MeV transverse 
momentum to the partons inside the proton (on the average). After a few 
such collisions the partons gain high enough transverse momenta to
become free of the proton and during this transition process they
do not interact (dead time). We assume that such a proton is ``lost''
for the reaction and does not participate in particle production
anymore. We note that such a picture corresponds to a modification
of the original Glauber model. It reduces the cross section for heavy
nuclei in central collisions and has no effect for light nuclei or
peripheral collisions. Furthermore, since central collisions have a small
weight in the total cross section, the value of the latter changes
insignificantly due to the proposed modification.

In technical terms the picture above corresponds to changing the
thickness expression
\be
\label{ABXm}
t_A(r) \,\, t_B(|\vec b - \vec r|)
\ee
in Eq.~(\ref{ABX}) in the following way: assuming that the 
nucleon dissociates after $N_c$ collisions, we divide the incoming ``rows''
into packets of $N_c$ nucleons. The first packet from the projectile collides 
with the first packet of the target and dissociates. This is followed by
colliding the next pair of packets and so on till the
last (incomplete) packets collide. Note that
unpaired packets will not produce any collisions in this framework. 
%since their counterparts are missing. 

Now we ask: how many collisions may the
proton suffer before it disintegrates (i.e. what is the value of $N_c$)?
We vary this parameter
to have the same $D/T$ ratio for all the three experiments studied.
The best fit is achieved with $N_c=4$ (right of Figure~\ref{figurenc}),
which, by random walk arguments, corresponds to a $\sqrt{N_c C}\approx$ 
1.25 GeV transverse momentum scale.

In the next subsection we study the plausibility of this idea in recent RHIC 
experiments.

\subsection{RHIC energy}

\begin{figure}
\epsfxsize=\textwidth \centerline{\epsffile{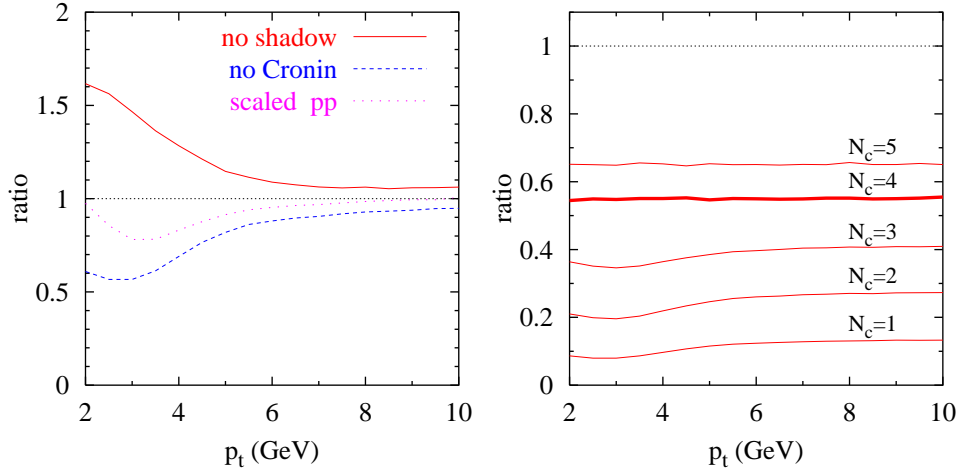}}
\vskip -0.05in
\caption[]{
 \label{figurerat}
Left: Ratio of invariant cross section neglecting different terms as compared
to the full pQCD calculation: neglecting shadowing (solid line), neglecting
Cronin effect (dashed line), and neglecting both (dotted line). Right:
Ratio of invariant cross section with proton dissociation compared to
the full pQCD calculation.
}
\end{figure}

Since recent RHIC experiments~\cite{RHIC} suggest a drastic reduction of the 
pion production cross section in central collisions, we now investigate
what effects may lead to such a suppression. One possibility is
jet quenching~\cite{jetq}, which takes into account the energy loss of partons
traveling through a diffractive medium. As a result, jets, normally
producing high $p_{\perp}$ mesons, are shifted to smaller transverse
momenta resulting in a large decrease of produced mesons at higher
tranverse momenta. The shift, or loss, may be dialed through the
properties of the surrounding matter (e.g. QGP). In order to be able to
assess those properties one has to know the uncertainties related to other
effects. Fig.~\ref{figurerat} (left) shows the influence of nuclear 
shadowing and of the Cronin effect in heavy ion collisions. Both of them have
a substantial impact modifying the pion production cross section by up to 
50\%, working in opposite directions. The uncertainty related
to them may render an assessment of jet quenching unreliable. The
Cronin effect was never studied systematically at higher energies; our
estimate is completely based
on a lower energy study ($\sqrt{s}\sim 30-40$ GeV).

Proton dissociation (studied in the previous subsection)
is another possible effect modifying theoretical predictions.   
We show suppressions produced by different
values of parameter $N_c$, indicating the number of collisions after
which the proton dissociates and does not participate in particle production.
Using the value deduced from the CERN SPS ($N_c=4$) reduces
the cross section by 45\% in Au+Au@130 AGeV collisions (indicated
by the thick line in the right panel of Fig. 3.). 
If we assume that at the higher energies of
RHIC the proton dissociation is more effective, than an assumption
of $N_c=3$ may be reasonable, resulting in a 60\% reduction. The proper
value should be extracted from a systematical $pA$ study planned at
RHIC and from the different centrality cuts. However, this value of
suppression is still too low to explain the experimental pion
production data in central collisions, leaving some room for
jet quenching.

\section{Conclusions}
In this talk we presented a pQCD based parton model augmented by the transverse
momentum distribution of the partons. The width of this distribution
is controlled by two terms, the $pp$ value, fixed by the experiments, and
a nuclear part, which gives extra enhancement due to semihard collisions.
We introduced the idea of proton dissociation, and
concluded on the basis of CERN SPS experiments that such an extension
of the Glauber model does not contradict the experiments. The best
value corresponds to 4 collisions before the proton
disintegrates, which is consistent with the picture of nuclear enhancement
of the transverse momentum distribution width obtained from
$pA$ collisions.

In high energy heavy ion collisions particle production at high
transverse momenta is a delicate interplay between intrinsic
transverse momentum enhancement, nuclear shadowing, the Cronin
effect, proton dissociation, and jet quenching. In order to
be able to separate all these effects one needs a systematic study
of $pp$, $pA$ and $AA$ reactions at different energies with the
same facility. RHIC with the planned $pA$ program provides a unique
opportunity to study the onset of the proton dissociation by 
increasing the target size and the onset of jet quenching in $AA$
collisions with centrality cuts or by the change of the projectile
size.

\section*{Acknowledgments}
This work was supported in part by  U.S. NSF grant INT-0000211,
DOE grant DE-FG02-86ER40251,  and Hungarian grants FKFP220/2000, OTKA-T032796 
and OTKA-T034842.

\end{document}